# Synthetic Voice Detection and Audio Splicing Detection using SE-Res2Net-Conformer Architecture

*Lei Wang\*, Benedict Yeoh, and Jun Wah Ng*

KLASS Engineering and Solutions Pte Ltd

{lei.wang, benedict.yeoh, junwah.ng}@klasses.com.sg

## Abstract

Synthetic voice and splicing audio clips have been generated to spoof Internet users and artificial intelligence (AI) technologies such as voice authentication. Existing research work treats spoofing countermeasures as a binary classification problem: *bonafide* vs. *spoof*. This paper extends the existing Res2Net by involving the recent Conformer block to further exploit the local patterns on acoustic features. Experimental results on ASVspoof 2019 database show that the proposed SE-Res2Net-Conformer architecture is able to improve the spoofing countermeasures performance for the logical access scenario.

In addition, this paper also proposes to re-formulate the existing audio splicing detection problem. Instead of identifying the complete splicing segments, it is more useful to detect the boundaries of the spliced segments. Moreover, a deep learning approach can be used to solve the problem, which is different from the previous signal processing techniques.

**Index Terms**: spoofing countermeasures, audio splicing detection, Res2Net, synthetic voice detection, deepfake

## 1. Introduction

While the rapid development of deep learning technologies enables the boom in AI applications, it also brings negative effects to the society and end users. One of the examples is the synthetic media, also known as deepfakes [1-5], which have been generated in the forms of audios [6-7], videos [5], images [4], etc. They could be potentially used to spread disinformation, impersonate individuals and commit crimes. Hence, the demand of synthetic media detection technologies becomes increasing [2].

Among the above-mentioned deepfakes, this work examines the detection methods for 2 types of fake audio data: synthetic voice and spliced audio. The synthetic voice is usually generated by machines and even state-of-the-art AI technologies such as Automatic Speaker Verification (ASV) is vulnerable to being manipulated by such data [6-8]. In the other approach of speech fraud, spliced audio clips have been produced to deceive Internet users [9-10].

Synthetic voice is commonly used to spoof ASV systems, and 2 types of techniques: text-to-speech (TTS) [11] and voice conversion (VC) [12] have been adopted to manipulate the speech. In spoofing countermeasures, such spoofing attack is known as Logical Access (LA) [6]; While the replay attack is another spoofing approach which is known as Physical Access, which is not covered in this paper. Although TTS and VC applications have been developed to benefit industrials such as multimedia and robotics, they have also been used in malicious ways to bring the society negative impacts such as fake news generation, person impersonation, telephone scams, etc. Hence, synthetic voice detection becomes an essential research topic in spoofing countermeasures.

Due to the rapid development of TTS and VC, one of the challenges in spoofing countermeasures is to detect fake speech generated by "unseen" techniques [13-17]. Hence it requires the spoofing countermeasures methods to be generalized and be able to predict those "unseen" techniques. The existing methods include deep learning structures such as light convolutional neural network (LCNN) [8], residual neural network (ResNet) [13], etc. that have been widely examined in image processing applications. The speech signal can be converted into time-frequency domain to extract representative feature vectors [18] which can be used as input of those deep neural networks. In recent studies [15-16, 19], Res2Net [20] with squeeze-and-excitation (SE) block has been shown to achieve higher generalizability to unseen spoofing attacks. It splits the feature block into multiple channel groups and residual connections are used across different groups. Hence, SE-Res2Net is capable to extract information from both local patterns and temporal patterns.

To further improve the performance of synthetic voice detection, this work attempts to exploit local and temporal patterns with more distinguishable power using other network structures. For example, the convolution-augmented transformer (Conformer) [21] combines convolution layers with self-attention module. Self-attention is capable to model global temporal context but it is less effective to capture local patterns. Convolution layers are good at extracting local information. Studies [21-22] show that Conformer can contribute to the speech-related tasks such as automatic speech recognition (ASR) and sound event detection. In this work, we propose to cascade the Conformer blocks after the SE-Res2Net to achieve better performance on spoofing detection task, and the new architecture is named SE-Res2Net-Conformer.

In addition, the synthetic voice detection problem is converted from a binary classification problem into a multi-classification problem which further separates "spoofing" data into "TTS" and "VC". It can benefit the end users by providing more precise information of how the spoofing speech has been generated. Moreover, the experimental results also show that slightly better performance of the system can be achieved.

Besides synthetic voice detection task, this paper also addresses the problem of audio splicing detection [9-10, 23-25]. Audio splicing becomes one of the most common ways to manipulate a target speaker's speech in the area of audio forensics [10]. The target speaker's speech at different times and in different places can be combined and edited into a new audio clip, however, the opinion of the speech content could be totally flipped. The existing splicing audio detection approaches [9-10, 25] usually rely on signal processing techniques to identify the spliced segments, e.g. reverberation and ambient noise characteristics. However, the spliced audio clips in real life can be more complicated and there are no clear indicators of original audio or spliced segments [1]. To deal with such scenario, this work proposes to detect only the splicing segment boundaries instead of detecting the complete splicing segments. It has 2 benefits: i) Evaluation metric becomes simpler; ii) Deep learning-based techniques can be easily applied.

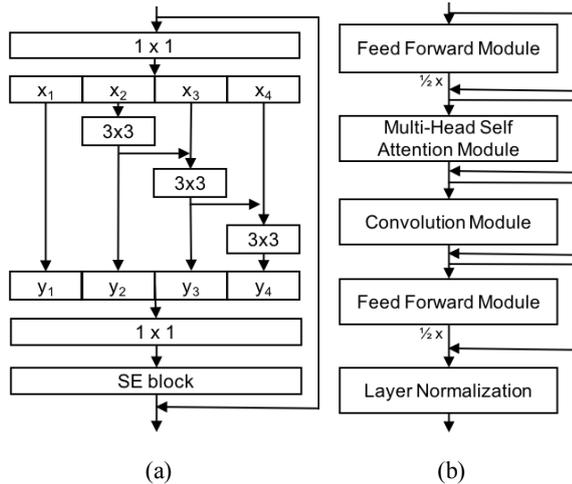

Figure 1: *(a) Illustration of a Res2Net block with a squeeze-and-excitation block; (b) Illustration of a Conformer block.*

The remaining of this paper is organized as followings: Section 2 introduces the proposed deep neural network architecture, and Section 3 presents the experimental results and discussion. We conclude in Section 4.

## 2. Model architecture

The proposed model architecture extends the existing Res2Net model with squeeze-and-excitation (SE) block [15] by appending self-attention Conformer blocks.

### 2.1. SE-Res2Net

Res2Net [20] has originally been proposed to extract multi-scale features in image processing. It extends the conventional bottleneck structure which is adopted in backbone convolution neural network (CNN) models. The 3×3 convolutional filters in the bottleneck block are replaced with smaller groups of filters, while connecting different filter groups in a hierarchical residual-like architecture, as shown in Figure 1(a).

Res2Net is able to extract a new dimension known as "multi-scale representation" which is a more effective factor in addition to depth, width and cardinality. It has been applied on object detection and semantic segmentation tasks [20]. Researchers in [15] have successfully applied Res2Net model on speech spoofing detection task, and its multi-scale representation contributes to predict unseen spoofing attacks.

Res2Net can also be integrated with other modern structures to further improve the model performance. For example, squeeze-and-excitation (SE) block has been cascaded to Res2Net to form a SE-Res2Net block, which has been shown benefit to anti-spoofing task. The SE block serves as a feature selection procedure to emphasize the most informative feature dimension.

### 2.2. Conformer

The Conformer [21] has been proposed to combine convolutional neural network and self-attention model hence to better handle both global temporal information and local feature patterns. Self-attention is capable to model global temporal context but it is less effective to capture local patterns. CNN is good at exploiting local information but many CNN layers need be cascaded to learn temporal information. Conformer has been shown to outperform the transformer-based model in ASR task.

The Conformer blocks have been used to replace the Transformer blocks in [26-27]. One Conformer block contains 2 half-step feed forward networks (FFNs) with a Multi-Head Self Attention Module (MHSA) and a Convolution Module in between. There are a few advantages: (i) The 2 half-step FFNs perform better than 1 single conventional FFN as studied in [28] which is known as Macaron-style; (ii) The attention-based module and convolutional module could perform at the same time hence to have the advantages of both convolutional network and self-attention network.

Besides adopting Conformer block to replace Transformer block, Conformer can also be used as a backbone model in classification tasks such as sound event detection task [22]. Conformer has been adopted to extract more distinguishable patterns from acoustic features hence to improve the F1 score of sound detection. It is believed that Conformer could also contribute to other classification problems. Hence, this work proposes to cascade the previous SE-Res2Net model together with the Conformer blocks hence to extract both global temporal and local patterns to benefit synthetic voice detection.

### 2.3. The cascaded SE-Res2Net-Conformer

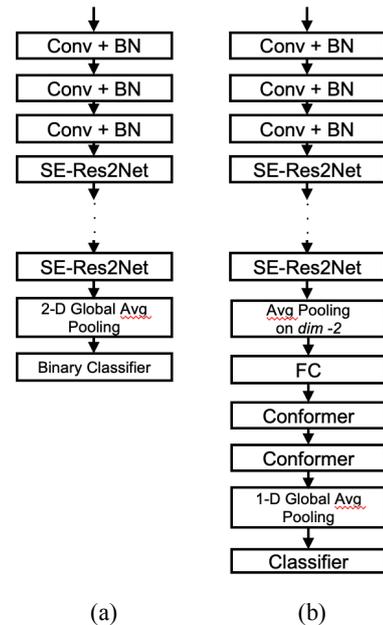

Figure 2: *(a) The baseline SE-Res2Net architecture used in [15]; (b) The proposed SE-Res2Net-Conformer architecture.*

Table 1: *No. of utterances of ASVspoof 2019 LA dataset.*

|  | **Bonafide** | **Spoof** | |
|---|---|---|---|
|  |  | **TTS** | **VC** |
| ***Train Set*** | 2,580 | 22,800 | |
|  |  | 15,200 | 7,600 |
| ***Dev Set*** | 2,548 | 22,296 | |
|  |  | 14,864 | 7,432 |
| ***Eval Set*** | 7,355 | 63,882 | |
|  |  | 49,140 | 14,742 |

The proposed SE-Res2Net-Conformer architecture is illustrated in Figure 2(b). It is an extension of SE-Res2Net which is shown in Figure 2(a). In the original SE-Res2Net architecture, the 3 convolution and batch normalization layers are used as encoder for the input speech features. A few SE-Res2Net blocks are then adopted to extract local patterns from the encoded features, and a 2-dimensional global average pooling is deployed before the classifier.

In the proposed SE-Res2Net-Conformer architecture, the global average pooling layer is replaced by an average pooling on the $2^{nd}$ last dimension. A full connection layer is then used followed by 2 Conformer blocks. A 1-dimensional global average pooling is then used before the classifier. The Conformer blocks are expected to extract more distinguishable information from the output of SE-Res2Net blocks.

## 3. Experimental setup and results

To verify the effectiveness of the proposed SE-Res2Net-Conformer architecture, it has been applied on both synthetic voice detection task and the splicing audio detection task.

### 3.1. Synthetic voice detection task

*3.1.1. Logical Access of ASVspoof 2019 Database*

The ASVspoof 2019 database [25] for LA contains both genuine speech and spoofed speech. The spoofed speech was generated using 17 unique spoofing algorithms including both TTS and VC. The database was split into 3 subsets: *train*, *dev* and *eval*. Table 1 shows the statistics of each subset.

The goal of the ASVspoof 2019 LA task is to classify the utterances into 2 categories: *bonafide* vs. *spoof*. In this work, the spoof utterances are further split into TTS and VC sub-categories according to the algorithms used. As the algorithms A13, A14 and A15 are combination of both TTS and VC, they are considered as machine synthesized speech so that the utterances generated using A13-15 are labelled as TTS.

*3.1.2. Acoustic Feature and Model Training*

In [15], acoustic features such as log power magnitude spectrogram, linear frequency cepstral coefficients, and constant-Q transform (CQT) [18] have been evaluated individually and CQT outperforms the other two. CQT has been proposed for music processing, and it employs geometrically spaced frequency bins. It is different from the Fourier transform-based features which impose regular spaced frequency bins. Hence, CQT has a balanced temporal resolution at both lower and higher frequencies.

In the experiment, CQT was extracted with 16ms step size using Hanning window, 9 octaves with 48 bins per octave, so that each CQT feature vector was 432 dimensional. The extracted CQT feature vectors were then truncated or repeated to form exactly 400 frames when input to the model.

In the experiment, both SE-Res2Net34 [15] and SE-Res2Net34-Conformer were examined. Moreover, it also did not detract from the objective of verifying the contribution of the Conformer. The models were implemented using TensorFlow 2.5. In SE-Res2Net34-Conformer, 2 Conformer blocks were used and each block was configured as follows: kernel size in the convolution module was 32, dropout rate was 0.2, feedforward layer factor for residual connection was 0.5, number of heads and head size of MHSA module were 32 and 16, respectively.

Table 2: *Results on the Eval Set of ASVspoof 2019 Logical Access. The EERs/t-DCFs of the 3-class models were generated in the same way as the 2-class scenario.*

|  | Sparse Categorical Accuracy (%) | EER(%) | t-DCF |
|---|---|---|---|
| SE-Res2Net50 (2-class) [15] | N.A. | 2.50 | 0.07 |
| Fusion result in [15] | N.A. | 1.89 | 0.05 |
| Our SE-Res2Net34 (2-class) | 90.42 | 4.79 | 0.14 |
| Our SE-Res2Net34 (3-class) | 87.92 | 2.08 | 0.06 |
| SE-Res2Net34-Conformer (3-class) | 96.86 | 1.85 | 0.06 |

The *train* set was used to train the model parameters while *dev* set was used to validate the model and guide the training process. The best model was selected based on the minimum loss value on *dev* set. The sparse categorical cross entropy loss function was adopted and performance metric was sparse categorical accuracy. Learning rate was set to 1e-3 with Adam optimization.

*3.1.3. Performance Metrics and Results*

To evaluate the performance of different models, 3 types of metrics are used: sparse categorical accuracy (SCA), equal error rate (EER) and t-DCF [29]. The SCA is commonly used to evaluate multi-classification models, especially when the classes are mutually exclusive. While EER is usually used in biometric security systems to predetermine a threshold value hence to balance false acceptance rate and false rejection rate. EER only deals with binary classification problem, e.g. *bonafide* vs. *spoof*. The t-DCF [29] has been proposed to evaluate ASV and anti-spoofing performance together. When EER and t-DCF are applied to evaluate the 3-class models, TTS and VC classes should be merged back to *spoof* class.

Table 2 shows the performance of different models on the *Eval* set. The SAC of the 3-class SE-Res2Net34 is 87.92% which is lower than the 2-class SE-Res2Net34 due to the misclassification between TTS and VC. However, the 3-class SE-Res2Net34 can reduce the EERs by relatively 56.6% and 16.8% comparing to the 2-class SE-Res2Net34 and the baseline SE-Res2Net50 [15], respectively. It shows that the model parameters estimated using 3 classes of data have better distinguishable power to separate *bonafide* speech from *spoof* speech. The proposed SE-Res2Net34-Conformer can further reduce the EER and t-DCF to 1.85% and 0.06, respectively. It outperforms the baseline SE-Res2Net50. When comparing to the feature fusion result in [15], the t-DCF of the proposed approach is worse but the EER is slightly better. In general, the improvement comes from the Conformer blocks which are able to exploit finer patterns from the acoustic features.

### 3.2. Splicing audio detection task

*3.2.1. Task description*

The existing work on splicing audio detection [9-10, 23-25] relies on the digital signal processing techniques to identify the locations of spliced segments. For example, acoustic reverberation and ambient noise have been analyzed to differentiate the spliced segments from the original audio [10, 25]. However, more complicated fake audio clips could be generated by combining multiple resources so that there are no obvious clues to indicate splicing or original segments. Thus, it is more important to know the transition boundaries where an audio segment from a different source is spliced.

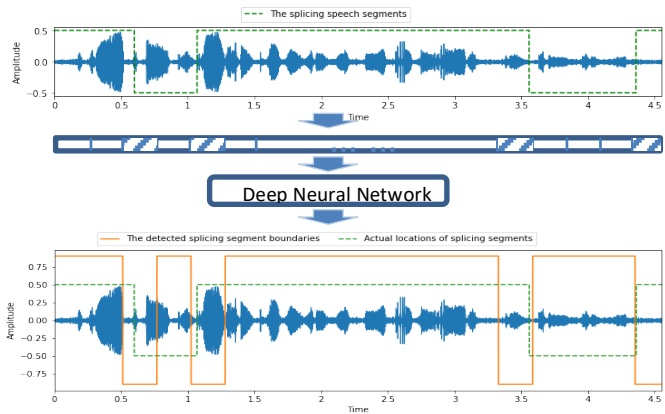

Figure 3: *The proposed framework to detect splicing segment boundaries. The green dotted line is the Boolean plot to indicate the portions of original speech (positive) vs. the spliced speech (negative). The shadowed chunks indicate that they contain splicing boundaries, i.e. the groundtruth. The orange solid line is the Boolean plot of the inferenced output, where the negative value indicates a splicing is detected. The scalar values of both plots have no specific meaning and they are chosen for better visualization purpose only.*

In this work, the splicing audio detection task aims to detect only the splicing segment boundaries instead of identifying the complete splicing segments. Inspired by speaker turn detection techniques [28], the brute force strategy is adopted to search for the splicing segment boundaries. As illustrated in Figure 3, a sliding window is employed to segment the audio clip into chunks, and acoustic features are extracted from each chunk. Deep learning approach can then be applied on each chunk of acoustic features to detect splicing segment boundaries.

The deep neural network is expected to extract local patterns from the chunk of acoustic features. The patterns are hypothesized to highlight the clues such as the discontinuity in speech, the inconsistency in the ambient noise, etc.

### 3.2.2. Spliced audio database and experimental setup

Due to lack of publicly available spliced audio database, the previous work created their own database for experiments. Inspired by [10], TIMIT [30] was chosen to generate the spliced database due to the availability of the word-level time stamps. Each spliced utterance was generated using a pair of utterances, $U_A$ and $U_B$, spoken by the same speaker: randomly selected 1 word from $U_B$ and inserted it into a random word boundary location of $U_A$. The process would be repeated for 1 – 3 times for each pair of utterances, i.e. each spliced utterance would have minimum 1 spliced word and maximum 3 spliced words. Table 3 summarizes the statistics of each set.

The studies [10, 25] show that ambient noise can be one of the clues to detect the spliced segments. It is keen to know whether ambient noise can also contribute to the deep learning approach. As TIMIT is a clean database, artificial noisy data noise was generated in this experiment. Air conditioner (aircon) noise was first added to the original TIMIT data at different SNR levels, and then the above-mentioned process was used to generate the noisy spliced audio database.

The CQT [18] was chosen to be the acoustic feature and its parameter settings followed Section 3.1.2. A block of CQT vectors were extracted from each chunk of the audio to input to the deep neural networks. The output was the class identifier which could be either "0" indicating "no splicing boundary" or "1" indicating "containing splicing boundary". In the implementation, the feature block size was 16 which was equivalent to 0.256 second.

In the experiment, the SE-Res2Net34 model that had about 0.96M weight parameters was evaluated on all types of spliced data. While the SE-Res2Net34-Conformer model was evaluated on the combination of different spliced data sets. The individual *train* set was insufficient to well estimate the model weights in SE-Res2Net34-Conformer which had more than 3.2M weight parameters. The model parameters and training configuration followed Section 3.1.2.

### 3.2.3. Experimental Results

The SCA is used to evaluate the model performance on the splicing boundary detection, and the evaluation results are presented in Table 4 and Table 5.

The SE-Res2Net model was applied on clean data, 25dB and 15dB noisy data, separately. The SCA on the clean data was the lowest, as shown in Table 4. It was expected as clean data had fewer acoustic clues to differentiate splicing segments from the original, e.g. lack of ambient noise. The SCA on the 15dB data was observed to be higher than it on 25dB, and it verified that the model captured the inconsistency in ambient noise. In Table 5, multi-conditional training was used to train both SE-Res2Net34 and SE-Res2Net34-Conformer models. The latter had better SCAs on all 3 *Eval* sets but the improvement was marginal. Its performance might be further improved by increasing the training data size.

## 4. Conclusions

This paper proposes the SE-Res2Net-Conformer architecture to exploit local and temporal patterns with more distinguishable power to improve the synthetic voice detection accuracy. In addition, the splicing audio detection problem is re-defined as to detect the spliced segment boundaries. It will allow deep learning-based techniques to be easily applied. Future work might include exploring other acoustic features such as log power magnitude spectrogram, studying the system performance on other noise types. In addition, pre-training on SE-Res2Net-Conformer could be examined to improve the model performance.

Table 3: *Statistics of the spliced TIMIT database.*

|  | No. Spliced Utterances | No. Speakers |
|---|---|---|
| **Train Set** | 23,030 | 462 |
| **Dev Set** | 2,493 | 50 |
| **Eval Set** | 5,888 | 118 |

Table 4: *The sparse categorical accuracy (%) on the Eval set of the spliced TIMIT database.*

|  | SE-Res2Net34 |
|---|---|
| Clean data | 83.79 |
| With aircon noise 25dB | 93.62 |
| With aircon noise 15dB | 94.36 |

Table 5: *Model performance comparison using multi-conditional training. The training data was the combination of the 3 Train sets used in Table 4. The clean Dev set was used to guide the training.*

| Sparse categorical accuracy (%) | Clean Eval | Aircon 25dB Eval | Aircon 15dB Eval |
|---|---|---|---|
| SE-Res2Net34 | 84.54 | 95.31 | 95.66 |
| SE-Res2Net34-Conformer | 85.12 | 95.33 | 95.73 |

# 5. References

[1] Video: "Image of the day: Donald Trump's confession speech that wasn't his," https://www.youtube.com/watch?v=tLHWCI3AxY0.

[2] "AI Singapore Trusted Media Challenge 2021," website: https://trustedmedia.aisingapore.org/.

[3] Tianxiang Chen, Avrosh Kumar, Parav Nagarsheth, Ganesh Sivaraman, and Elie Khoury, "Generalization of Audio Deepfake Detection," in *Proc. Odyssey 2020 The Speaker and Language Recognition Workshop*, Tokyo, Japan, Nov. 2020, pp. 132-137.

[4] Sheng Li, Xun Zhu, Guorui Feng, Xinpeng Zhang, and Zhenxing Qian, "Diffusing the Liveness Cues for Face Anti-spoofing," in *Proceedings of the 29th ACM International Conference on Multimedia (MM '21)*, Virtual Event, China, Oct. 2021, pp. 1636-1644.

[5] Trisha Mittal, Uttaran Bhattacharya, Rohan Chandra, Aniket Bera, and Dinesh Manocha, "Emotions Don't Lie: An Audio-Visual Deepfake Detection Method using Affective Cues," in *Proceedings of the 28th ACM International Conference on Multimedia (MM'20)*, Seattle, WA, USA, Oct. 2020, pp. 2823–2832.

[6] Andreas Nautsch, Xin Wang, Nicholas Evans, Tomi Kinnunen, Ville Vestman, Massimiliano Todisco, Hector Delgado, Md Sahidullah, Junichi yamagishi, and Kong Aik Lee, "ASVspoof 2019: spoofing countermeasures for the detection of synthesized, converted and replayed speech," in *IEEE Trans. on Biometrics, Behavior, and Identity Science*, vol.3, no.2, Apr. 2021, pp. 252-265.

[7] Run Wang, Felix Juefei-Xu, Yihao Huang, Qing Guo, Xiaofei Xie, Lei Ma, and Yang Liu, "DeepSonar: Towards Effective and Robust Detection of AI-Synthesized Fake Voices," in *Proceedings of the 28th ACM International Conference on Multimedia (MM '20)*, Seattle, WA, USA, Oct. 2020, pp. 1207-1216.

[8] Zhenzong Wu, Rohan Kumar Das, Jichen Yang, and Haizhou Li, "Light Convolutional Neural Network with Feature Genuinization for Detection of Synthetic Speech Attacks," in *Proc. INTERSPEECH 2020*, Shanghai, China, Oct. 2020, pp. 1101-1105.

[9] Qi Yan, Rui Yang, and Jiwu Huang, "Robust copy-move detection of speech recording using similarities of pitch and formant," in *IEEE Transactions on Information Forensics and Security*, vol.14, no.9, Sept. 2019, pp. 2331-2341.

[10] Hong Zhao, Yifan Chen, Rui Wang, and Hafiz Malik, "Audio splicing detection and localization using environmental signature," in *Multimedia Tools and Applications*, 2017, pp. 13897-13927.

[11] Xu Tan, Tao Qin, Frank Soong, and Tie-Yan Liu, "A Survey on Neural Speech Synthesis," arXiv preprint, 2021. https://arxiv.org/abs/2106.15561.

[12] Berrak Sisman, Junichi Yamagishi, Simon King, and Haizhou Li, "An overview of voice conversion and its challenges: from statistical modeling to deep learning," in *IEEE/ACM Transactions on Audio, Speech, and Language Processing*, vol.29, Nov. 2020, pp. 132-157.

[13] Cheng-I Lai, Nanxin Chen, Jesús Villalba, and Najim Dehak, "ASSERT: Anti-Spoofing with Squeeze-Excitation and Residual neTworks," in *Proc. INTERSPEECH 2019*, Graz, Austria, Sept. 2019, pp. 1013-1017.

[14] Galina Lavrentyeva, Sergey Novoselov, Andzhukaev Tseren, Marina Volkova, Artem Gorlanov, and Alexandr Kozlov, "STC Antispoofing Systems for the ASVspoof2019 Challenge," in *Proc. INTERSPEECH 2019*, Graz, Austria, Sept. 2019, pp. 1033-1037.

[15] Xu Li, Na Li, Chao Weng, Xunying Liu, Dan Su, Dong Yu, and Helen Meng, "Replay and Synthetic Speech Detection with Res2Net Architecture," in *Proceedings of ICASSP 2021*, Toronto, Canada, June 2021.

[16] Xu Li, Xixin Wu, Hui Lu, Xunying Liu, and Helen Meng, "Channel-Wise Gated Res2Net: Towards Robust Detection of Synthetic Speech Attacks," in *Proc. INTERSPEECH 2021*, Brno, Czechia, Aug./Sept. 2021, pp. 4314-4318.

[17] Yexin Yang, Hongji Wang, Heinrich Dinkel, Zhengyang Chen, Shuai Wang, Yanmin Qian, and Kai Yu, "The SJTU Robust Anti-Spoofing System for the ASVspoof 2019 Challenge," in *Proc. INTERSPEECH 2019*, Graz, Austria, Sept. 2019, pp. 1038-1042.

[18] Massimiliano Todisco, Héctor Delgado, and Nicholas Evans, "Constant Q Cepstral Coefficients: A Spoofing Countermeasure for Automatic Speaker Verification," in *Computer Speech & Language*, vol.45, Feb. 2017, pp.516-535.

[19] Xin Fang, Haijia Du, Tian Gao, Liang Zou, and Zhenhua Ling, "Voice spoofing detection with raw waveform based on Dual Path Res2net," in *Proceedings of the 5th International Conference on Crowd Science and Engineering (ICCSE '21)*, Jinan, China, Oct. 2021, pp.160-165.

[20] Shang-Hua Gao, Ming-Ming Cheng, Kai Zhao, Xin-Yu Zhang, Ming-Hsuan Yang, and Philip Torr, "Res2Net: A new multi-scale backbone architecture," in *IEEE Trans. on Pattern Analysis and Machine Intelligence*, vol.43, no.2, Feb. 2021.

[21] Anmol Gulati, James Qin, Chung-Cheng Chiu, Niki Parmar, Yu Zhang, Jiahui Yu, Wei Han, Shibo Wang, Zhengdong Zhang, Yonghui Wu, and Ruoming Pang, "Conformer: Convolution-augmented Transformer for Speech Recognition," in *Proc. INTERSPEECH 2020*, Shanghai, China, Oct. 2020, pp. 5036-5040.

[22] Koichi Miyazaki, Tatsuya Komatsu, Tomoki Hayashi, Shinji Watanabe, Tomoki Toda, and Kazuya Takeda, "Convolution-augmented transformer for semi-supervised sound event detection," in *Proceedings of the Workshop on Detection and Classification of Acoustic Scenes and Events 2020*, Japan, Nov. 2020.

[23] Muhammad Imran, Zulfiqar Ali, Sheikh Tahir Bakhsh, and Sheeraz Akram, "Blind Detection of Copy-Move Forgery in Digital Audio Forensics," in *IEEE Access*, vol.5, June 2017, pp. 12843-12855.

[24] Kasi Mannepalli, Pelluri Krishna, Kovi Krishna, and Kodali Krishna, "Copy and Move Detection in Audio Recordings using Dynamic Time Warping Algorithm," in *International Journal of Innovative Technology and Exploring Engineering*, vol.9, no.2, Dec. 2019, pp. 2244-2249.

[25] Xunyu Pan, Xing Zhang, and Siwei Lyu, "Detecting splicing in digital audios using local noise level estimation," in *Proceedings of ICASSP 2012*, Kyoto, Japan, March 2012.

[26] S. Karita *et al.*, "A comparative study on transformer vs RNN in speech applications," in *Proceedings of ASRU 2019*, Singapore, Dec. 2019.

[27] Qian Zhang, Han Lu, Hasim Sak, Anshuman Tripathi, Erik McDermott, Stephen Koo, and Shankar Kumar, "Transformer transducer: A streamable speech recognition model with transformer encoders and RNN-T loss," in *Proceedings of ICASSP 2020*, Spain, May 2020.

[28] Margarita Kotti, Emmanouil Benetos, and Constantine Kotropoulos, "Computationally Efficient and Robust BIC-Based Speaker Segmentation," in *IEEE Trans. Audio Speech Language Processing*, vol.16, no.5, July 2008.

[29] Tomi Kinnunen, Héctor Delgado, Nicholas Evans, Kong Aik Lee, Ville Vestman, Andreas Nautsch, Massimiliano Todisco, Xin Wang, Md Sahidullah, Junichi Yamagishi, and Douglas A. Reynolds, "Tandem assessment of spoofing countermeasures and automatic speaker verification: Fundamentals," in *IEEE ACM Trans. Audio Speech Language Processing*, vol. 28, July 2020, pp. 2195–2210.

[30] John S. Garofolo, Lori F. Lamel, William M. Fisher, Jonathan G. Fiscus, David S. Pallett, Nancy L. Dahlgren, and Victor Zue, "TIMIT Acoustic-Phonetic Continuous Speech Corpus LDC93S1," Web Download. Philadelphia: Linguistic Data Consortium.